\documentclass[aps,prb,amsmath,twocolumn,amssymb,floatfixng,showpacs,
superscriptaddress,footinbib,multicol]{revtex4-1}
\usepackage[dvips]{graphics}
\usepackage{bm}
\usepackage{float}
\usepackage{epsfig}
\usepackage{enumerate}
\usepackage{amsmath}
\usepackage{color}
\usepackage{graphicx}
\usepackage[colorlinks=true,linktoc=page,linkcolor=red,citecolor=blue]{hyperref}

\newcommand{\ie}{{\it i.e.,}}
\newcommand{\eg}{{\it e.g.}}

\newcommand\bea{\begin{eqnarray}}
\newcommand\eea{\end{eqnarray}}
\newcommand\beq{\begin{equation}}  
\newcommand\eeq{\end{equation}}

\begin{document} 
\title{Tailoring Metal Insulator Transitions $\&$ Band Topology via Off-resonant Periodic Drive in an Interacting Triangular Lattice} 
\author{Sayan Jana}
\email{sayan@iopb.res.in}
\affiliation{Institute of Physics, Sachivalaya Marg, Bhubaneswar-751005, India}
\affiliation{Homi Bhabha National Institute, Training School Complex, Anushakti Nagar, Mumbai 400085, India}
\author{Priyanka Mohan}
\email{priyankamohan@theory.tifr.res.in}
\affiliation{Department of Theoretical Physics, Tata Institute of Fundamental Research, Homi Bhabha Road, Mumbai 400005, India}
\author{Arijit Saha}
\email{arijit@iopb.res.in}
\affiliation{Institute of Physics, Sachivalaya Marg, Bhubaneswar-751005, India}
\affiliation{Homi Bhabha National Institute, Training School Complex, Anushakti Nagar, Mumbai 400085, India}
\author{Anamitra Mukherjee}
\email{anamitra@niser.ac.in}
\affiliation{School of Physical Sciences, National Institute of Science Education and Research, HBNI, Jatni 752050, India}

\begin{abstract}
A triangular lattice with onsite Coulomb interaction $U$ present only on one sub-lattice, is periodically driven by electromagnetic field with a frequency $\Omega \gg (t,~U)$ at half filling. In this high frequency limit, the electromagnetic vector potential, with an amplitude $A$, modifies the bare hopping and generates new next nearest neighbour hopping parameters. For $U=0$, the driving acts like an emergent intrinsic spin-orbit coupling term and stabilises three dispersive bands with the lower and upper bands having non zero Chern numbers. Within a slave rotor mean field theory, we show that while $U$ freezes out charge fluctuations on the interacting sub-lattice, it does not open up a charge gap without the external drive. In presence of the drive, and small $U$, the system exhibits repeated metal insulator transitions as a function of the amplitude $A$. For large $U$, we establish that the freezing of charge fluctuations on the interacting sub-lattice stabilizes an emergent, low energy \textit{half filled non-interacting Kane-Mele model}, whose band gaps can be tuned by varying $A$. In this limit, we show that the external drive provides an handle to engineer periodic band inversions at specific values of $A$ accompanied by topological phase transitions that are characterised by swapping of band Chern numbers. 
\end{abstract}

\maketitle
\section{Introduction}
In the last decade, the advent of topological insulators (TIs) has caused to a revolutionary impact on the concept of band structure of materials, both on the  theoretical~\cite{kane2005quantum, bernevig2006quantum, moore2010birth, qi2011topological, annreview2DTI, annreview3DTI, TKNNformula, fukaneTI2007, teofukaneTI2008, pesinmacdonald2012, nayak2008non, ezawa2012topological, quancompastern2013, kane2005quantumZ2} as well as on the experimental~\cite{konig2007quantum,culcer2012transport, xia2009observation, zhang2009topological, hasan2010colloquium} front. This has lead to intense theoretical~\cite{rachel2018interacting,rachel2010topological,raghu2008topological,pesin2010mott,dzero2010topological,tran2012phase,sayan2019lieb} as well as experimental~\cite{shitade2009quantum,chadov2010tunable,lin2010half} 
investigation of the effects of strong correlation in systems that host nontrivial topological bands. 
Very recently, there has been an upsurge of research activity in emergent topological phases in \textit{out of equilibrium} non-topological systems via external periodic driving~\cite{oka2009photovoltaic,lindner2011floquet,dora2012optically}. 
Experimental feasibility of engineering such periodically driven systems~\cite{wang2013observation, rechtsman2013photonic, jotzu2014experimental, peng2016experimental} has opened up the opportunity 
for investigating existence of non equilibrium Majorana modes~\cite{jiang2011majorana, kundu2013transport, perez2014floquet}, non trivial transport properties~\cite{gu2011floquet, kundu2014effective, farrell2015photon} as well as controlling of band structure~\cite{sambe1973steady, klinovaja2016topological}, disorder effect~\cite{titum2015disorder}. 
The effect of drive on the tight binding band structure in graphene~\cite{kitagawa2011transport, usaj2014irradiated},
spin-orbit coupled Dirac materials like silicene and  germanene~\cite{ezawa2013photoinduced, drummond2012electrically, liu2011quantum, mohan2016brillouin} and low energy spectrum of semi-Dirac materials~\cite{PhysRevB.94.081103, PhysRevB.98.235424}, are being actively pursued.

Given this background, it is natural to investigate the interplay between strong local electronic correlations and external electromagnetic driving on different model systems. This is a rather broad question and some aspects have been addressed in recent literature. There are three  regimes where different theoretical tools can be applied, small driving frequency $\hbar\Omega \ll U$, near resonance $\hbar\Omega\sim U$ and off resonance $\hbar\Omega \gg U$. Dynamical localization has been investigated in one dimensional spin half Fermi systems~\cite{PhysRevB.95.014305} within small frequency approximation. Effective spin model for one and two orbital Hubbard model at half filling has been studied~\cite{PhysRevB.99.205111}. The effect of near and off resonant driving on the double occupation in the Mott state has been investigated~\cite{PhysRevLett.121.233603}. Also perturbative analysis of driving on Kondo insulators~\cite{PhysRevB.96.115120, PhysRevX.4.041048} has been undertaken. The Bose Hubbard model has been studied in the off resonant (high frequency) regime both experimentally~\cite{PhysRevLett.115.155301} and theoretically~\cite{PhysRevB.92.125107,PhysRevLett.109.203005,PhysRevLett.103.133002}.

While these have added valuable insights to the physics of driven systems, explicit study of metal insulator transition (MIT) under driving, for Fermi systems, has not yet been addressed. Even more unexplored 
is the nature of MIT, when the underlying bands are driven into a topologically non trivial regime. Here we investigate these issues for a triangular lattice with diluted Hubbard interaction at half filling. One of the concerns of periodic driving of an isolated many body system is that, it leads to a featureless thermal state in long time limit~\cite{PhysRevX.4.041048}. However, it has been shown that the time scale of heating is exponentially/quasi exponentially slow~\cite{PhysRevLett.115.256803,Abanin2017,PhysRevE.90.012110} allowing one to work in a pre-thermal regime which survives for experimentally relevant time scales~\cite{heating}. With this justification in our study we incorporate the drive via Peierls's substitution to the bare tight binding hopping through a time dependent vector potential $\mathcal{A}(t)=\{A_x \cos{\Omega t}, A_y\cos{(\Omega t -\phi})\}$. We denote the amplitude of the vector potential by $A$. We employ high frequency Brillouin-Wigner perturbation theory~\cite{bw} to obtain a quasi-static effective Hamiltonian ($H_{eff}$) in the zero photon subspace upto the order of 1/$\Omega$. The effect of driving gives rise to correction to the nearest neighbour (NN) bare hopping elements between different sub-lattices as well as generates new next-nearest neighbour (NNN) hopping terms~\cite{ofr}. These emergent hopping amplitudes are chiral in nature and act like intrinsic spin-orbit coupling, which leads to topologically non trivial bands in the Floquet quasi-energy spectrum~\cite{quasi}. While magnetism is certainly an important feature of the Hubbard model, here our focus lies in the study of the effect of drive only on charge fluctuations. Hence, we study our model in the paramagnetic regime within a slave rotor mean field theory (SR-MFT)~\cite{florens2004slave, zhao2007self}. 
\begin{figure}[t]
	\centering{
		\includegraphics[width=8.0cm, height=4.5cm, clip=true]{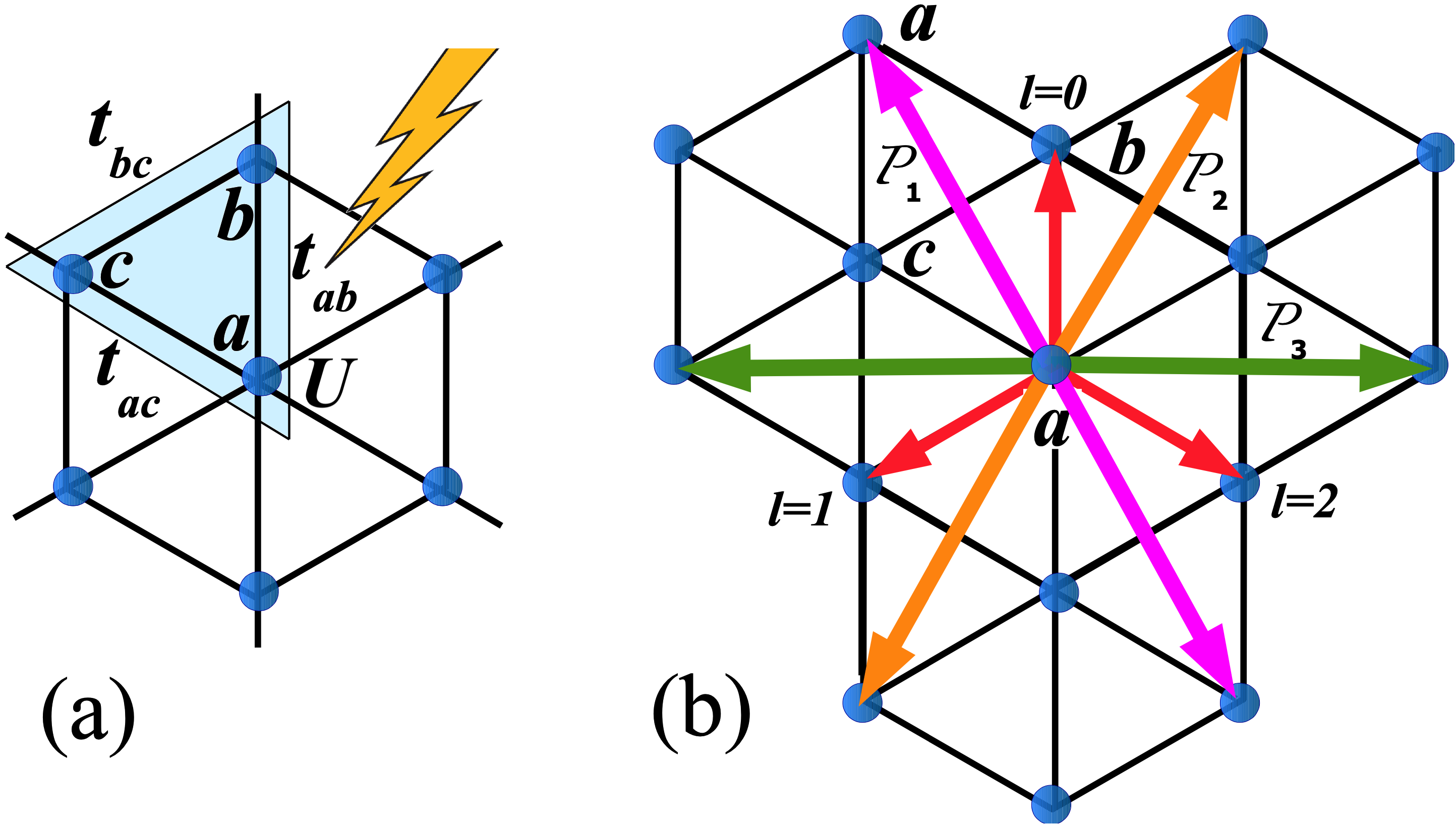}}
	\caption{(Color online) 
	(a) Schematic structure of the triangular lattice is demonstrated. Each unit cell consists of three atoms $a$, $b$ and $c$ shown enclosed in a triangle. Here, $t_{ab}$, $t_{bc}$, $t_{ca}$
             represent the nearest neighbour hopping amplitudes between the three atoms. $U$ denotes the strength of onsite Hubbard interaction on the $a$ sublattice. The lightning bolt represents the external electromagnetic radiation. (b) The thin red arrows indicate the three directional vectors $l=0,1,2$ in the triangular lattice. The thick, magenta, orange and green arrows denote the light induced next nearest neighbour hopping along the $\text{path-1} (\mathcal{P}_{1})$, $\text{path-2} (\mathcal{P}_{2})$ and $\text{path-3} (\mathcal{P}_{3})$ respectively within $a$ sublattice. Similar terms also exist for both $b$ and $c$ sublattices. The chirality of light induced hopping 
terms are discussed in the text.}
	\label{fig-1}
\end{figure}

In the uncorrelated model with three dispersive bands, non-zero $A$ generates chiral NNN hopping terms and also modifies the bare hopping elements. The resulting Hamiltonian hosts topologically non-trivial lower and upper bands and a trivial dispersive middle band. Variation of $A$ causes periodic touching of the three bands. Switching on $U$ on one sub-lattice (on one site of the three site unit cell of the triangular lattice), induces a `local Mott transition' beyond a threshold value of the correlation strength $U_{Crit}$ that  suppresses charge fluctuations at the correlated sub-lattice. In presence of the drive, and for $U>U_{Crit}$, we show that the local Mott transition, the drive modulated NN hopping and the emergent NNN hopping terms, conspire to stabilize an insulating state. This insulating state is characterized by topologically non-trivial low energy bands split by a small charge gap and high energy bands separated by $U$. We show that, similar to the $U=0$ case, the charge gap between the low energy bands oscillate periodically with $A$ and stabilizes a semi-metal at specific values of $A$ where the charge gap goes to zero. Moreover at each such band touching there is a topological phase transition whereby the Chern number of the topologically non-trivial bands are exchanged. For a range of bare hopping parameters, we establish that this periodic gap closing and swapping of Chern numbers, holds for the entire insulating regime and is independent of $U$, once $U>U_{Crit}$. We establish that this phenomenology can be understood in terms of an emergent low energy Hamiltonian that resembles \textit{the half filled non-interacting Kane-Mele model}. Further, based on single particle density of states, we map out $U_{Crit}$ as a function of $A$ and present the $U$ vs $A$ phase diagram, emphasizing how the peaks and dips of $U_{Crit}$ are governed by a competition between the interaction strength and the drive induced bandwidth modulations.

The remainder of the paper is organized as follows. In Sec.~\ref{sec:II} we discuss our model, the slave rotor mean-field theory, and the observables. In Sec.~\ref{sec:III} we present our numerical results. Finally, we summarize and conclude the paper in Sec.~\ref{sec:IV}.

\section{Model and method \label{sec:II}}
We consider a triangular lattice as our model system with $a$, $b$ and $c$ sub-lattices as shown in Fig.~\ref{fig-1}(a). The Hamiltonian of the system is defined as $H=H_{Free}+H_{Int}$. The tight binding Hamiltonian $H_{Free}$ can be written as
\begin{eqnarray}
H_{Free}&=&H_{ab}+H_{bc}+H_{ac}\nonumber\\
&=&-\sum_{\langle i, j\rangle} (t_{ab} a^\dagger_i b_j+t_{bc} b^\dagger_i c_j+ t_{ca} a^\dagger_i c_j +h.c.) \ ,
\label{Eq1}
\label{triangular}
 \end{eqnarray}
where, $t_{ab}$, $t_{bc}$ and $t_{ac}$ represent the nearest-neighbour (NN) hopping amplitudes between the $a, b$ and $c$ sub-lattices. Here $a^\dagger_i (a_{i}), b^\dagger_i (b_{i})$ and $c^\dagger_i (c_{i})$ correspond to the creation (annihilation) operators for the $a, b$ and $c$ sub-lattices respectively. For notational clarity, the spin indices are suppressed for now. The main steps here consists of first treating the effect of driving on the $H_{Free}$ and then study the effect of interaction on the emergent tight binding model. This is the standard procedure in the high frequency limit~\cite{PhysRevLett.121.233603}. 

\textit{\underline{The effective kinetic energy Hamiltonian}:} In our analysis, we consider the high-frequency limit of the periodic driving. In this limit, one can derive an effective quasi-static Hamiltonian taking into account only the virtual photon transitions. Technically, the full Floquet Hamiltonian in the extended Sambe space is projected back to the zero photon subspace using a high-frequency expansion based on the Brillouin-Wigner (BW) perturbation theory~\cite{mikami2016brillouin}. 
In comparison to Floquet-Magnus~\cite{mananga2011introduction,blanes2009magnus} and van Vleck expansions~\cite{eckardt2015high,bukov2015universal}, higher order terms are easier to calculate using BW expansion.

The vector potential of the electromagnetic radiation is given by, 
 \begin{eqnarray}
 \mathcal{A}(t)=(A_x \cos{\Omega t}, A_y\cos{(\Omega t -\phi)})\ ,
 \label{vecpot}
 \end{eqnarray}
where $\Omega$ is the frequency of the light, $A_{x}, A_{y}$ are the amplitudes of the irradiation and $\phi$ is the phase. The polarization of the laser can be controlled by choosing appropriate 
values for $A_x/A_y$ and $\phi$. For \eg, circular polarization can be obtained by choosing $A_x=A_y$, $\phi=\pi/2$ and linear polarization by $\phi=0$ respectively. 
The other values of $A_x/A_y$ and $\phi$  correspond to elliptical polarized light.
 
The vector potential of the external irradiation is incorporated by Peierl’s substitution thereby transforming the above Hamiltonian time-dependent. The hopping elements thus acquire a phase given by
\begin{eqnarray}
t\rightarrow t e^{-i(r_1 \sin{\Omega t}+r_2 \cos{\Omega t})}\ ,
\end{eqnarray}
where $r_1(l)={\tilde{a}} A_y \sin(\phi) \cos(\frac{2\pi l}{3})$ and $r_2(l)={\tilde{a}}\left[A_y \cos(\phi) \cos(\frac{2\pi l}{3})-A_x \sin(\frac{2\pi l}{3}) \right]$. Here  $l=0,1,2$ denotes the three directions 
 within the lattice as shown in Fig.~\ref{fig-1}(b) and $\tilde{a}$ is the lattice spacing.

The  Floquet Hamiltonian~\cite{shirley1965solution,sambe1973steady} can be defined in the following way:
\begin{eqnarray}
 H^p=\frac{1}{T}\int^T_0 H(t) e^{ip\Omega t} d t\ ,
 \label{Hfloquet}
\end{eqnarray}
As discussed in the Appendix~\ref{sec: appendix A}, the final form of the Floquet Hamiltonian is 

\begin{eqnarray}
 H_{K}=H^0+ \sum_{n\ne 0}\frac{H^{-n} H^{n}}{n\Omega}\ ,
 \label{bw}
 \end{eqnarray}
For the triangular lattice under our consideration (see Fig.~\ref{fig-1}(a)), the zeroth order term is given by,
\begin{eqnarray}
 H^0&=& -\sum_{\langle ij \rangle} J_{0}\left(\sqrt{r^2_1+r^2_2}\right) \nonumber\\
 &&\left(t_{ab}\,  a^\dagger_i b_j + t_{bc}\,  b^\dagger_i c_j+t_{ca}\,  a^\dagger_i c_j +h.c.\right).
 \label{zeroorder}
\end{eqnarray}
In the zeroth order Hamiltonian, the effects due to periodic driving are manifested in renormalized hopping parameters. 
The first order term from Eq.(\ref{bw}) results in an effective spin-orbit coupling as described later. Similar terms has also been reported
earlier in case of hexagonal lattice (graphene)~\cite{mikami2016brillouin}, spin-orbit coupled Dirac materials~\cite{mohan2016brillouin}.  
In our case, the $O(t^2/\Omega)$ term is shown in Eq.(\ref{m-so}). 

In Eq.(\ref{m-so}), $\nu_{ij}=\pm 1$  depending on whether the next nearest neighbour (NNN) hopping takes place in clockwise or anti-clockwise manner. This introduces an intrinsic chirality in the model, similar to intrinsic spin orbit coupling~\cite{kane2005quantum}. In Fig.~\ref{fig-1}(b), this external light induced spin-orbit coupling terms are denoted by paths ${\mathcal{P}}_{1}, {\mathcal{P}}_{2}$ and ${\mathcal{P}}_{3}$ within the $a$ sub-lattice. Similar terms are also present for $b$ and $c$ sub-lattices which can be seen from Eq.(\ref{m-so}). The full analytical form of $\chi_{1,2,3}$ in Eq.(\ref{m-so}) is provided in the Appendix~\ref{sec: appendix A}, where it is shown that for circularly polarized light,  $\chi_{1}=\chi_2=\chi_{3}$. Also, from Eq.(\ref{m-so}) it is apparent that for $t_{ab}=t_{bc}=t_{ca}$, the $1/\Omega$ order term vanishes. 
\begin{widetext}
\begin{eqnarray}
 \frac{H^{-n} H^{n}}{n\Omega}&=&
 \sum^{{\mathcal{P}}_{1}}_{\langle \langle ij \rangle\rangle} \chi_1 \, \nu_{ij} \, \, \Big[ (t^2_{ab}-t^2_{ca})   a^\dagger_i a_{j} 
+(t^2_{ab}-t^2_{bc})  b^\dagger_i b_{j}  +(t^2_{ac}-t^2_{bc})  c^\dagger_i c_{j} \Big] \nonumber\\
&+&\sum^{{\mathcal{P}}_{2}}_{\langle \langle ij \rangle\rangle} \chi_2 \, \nu_{ij} \, \,\Big[ (t^2_{ab}-t^2_{ca})   a^\dagger_i a_{j} 
+(t^2_{ab}-t^2_{bc})  b^\dagger_i b_{j}  +(t^2_{ac}-t^2_{bc})  c^\dagger_i c_{j} \Big]  \nonumber\\
 &+&\sum^{{\mathcal{P}}_{3}}_{\langle \langle ij \rangle\rangle}\chi_3 \, \nu_{ij} \,  \, \Big[ (t^2_{ab}-t^2_{ca})   a^\dagger_i a_{j} 
+(t^2_{ab}-t^2_{bc})  b^\dagger_i b_{j}  +(t^2_{ac}-t^2_{bc})  c^\dagger_i c_{j} \Big].
\label{m-so}
\end{eqnarray}
\end{widetext}
This is different from honeycomb lattices where the light induced spin-orbit term is non-zero irrespective of the polarization of the external irradiation~\cite{mikami2016brillouin}. In triangular lattices, the spin-orbit coupling term between each pair of next-neighbour sites has two contributions with opposite chirality. Thus, they can nullify each other if the hopping amplitudes are equal to each other. Here, we define
the light induced NNN hopping amplitudes as $t_{aa}=\chi_{1}\nu_{ij}(t_{ab}^{2} - t_{ca}^{2})$, $t_{bb}=\chi_{2}\nu_{ij}(t_{ab}^{2} - t_{bc}^{2})$, $t_{cc}=\chi_{3}\nu_{ij}(t_{ac}^{2} - t_{bc}^{2})$.

\textit{\underline{Interaction effects}:} 
To discuss the effects of $H_{Int}$, we first rewrite $H_K$ is a succinct notation as a tight binding model on a triangular lattice with nearest and chiral NNN hopping and also make the spin indices explicit. The unit cell consists of three atoms indicated by $a$, $b$ and $c$ as shown in Fig.~\ref{fig-1}(a). Hence, the kinetic term ($H_{K}$) in a compact form can be written as,
\begin{eqnarray}\label{e3}
&& H_{K}=\sum_{ 
I,J,\alpha,\beta,\sigma}(t^{A}_{I\alpha\sigma;J\beta\sigma}
d_{I\alpha\sigma}^{\dagger}d_{J\beta \sigma} + 
h.c.)\ ,
\end{eqnarray}  
where the indices $I, J$ denote the three site unit cells of the lattice. The hopping terms depend on the strength $A$ of the external periodic drive. 
Here, $I$ equal to $J$ implies the hopping within a unit cell, while $I\neq J$ indicates to the hopping between different unit cells, while $\alpha$ and $\beta$ runs over the three atomic labels $a$, $b$ and $c$. $t^{A}_{I\alpha\sigma;J\beta\sigma}$ parameters are appropriately chosen to incorporate NN and NNN hopping elements as necessary. 
The Hubbard interaction is introduced only on the $a$ sub-lattice. 
The final form of the off-resonant Hamiltonian\cite{ofr} reads as: 
\begin{eqnarray}\label{e4}
H^{\rm OR}_{\rm eff} 
&=&\sum_{ 
I,J,\alpha,\beta,\sigma}(t^{A}_{I\alpha\sigma;J\beta\sigma}
d_{I\alpha\sigma}^{\dagger}d_{J\beta \sigma}+h.c.)\nonumber\\&+&\sum_{I}Un_{Ia\uparrow}n_{Ia\downarrow}
\end{eqnarray}  
\begin{figure*}[t]
\centering{
\includegraphics[width=16cm, height=8cm, clip=true]{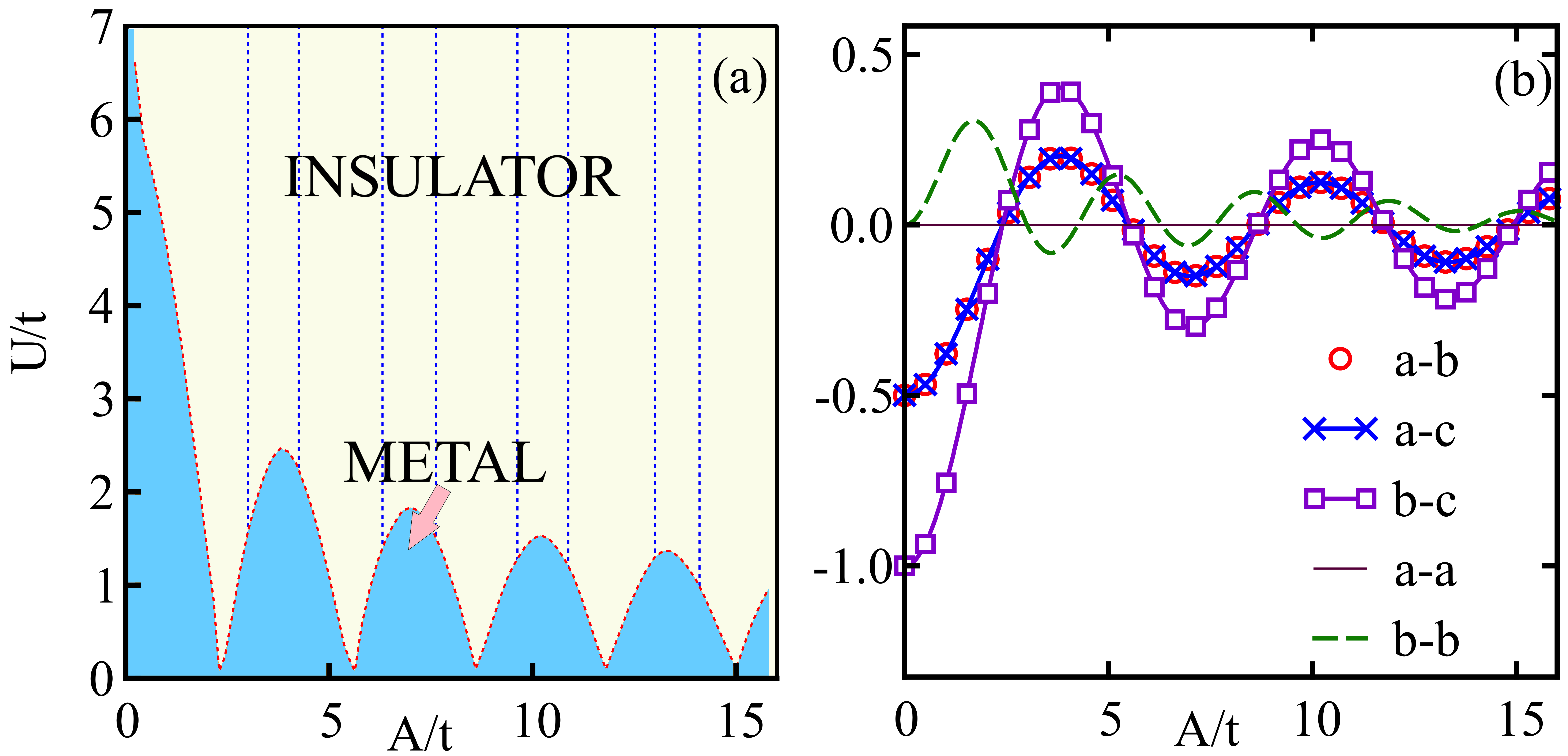}}
\caption{(Color online) The $U-A$ phase diagram is shown in panel (a). The red (dashed) line with squares depicts the evolution of $U_{Crit}$ with $A/t$. The region below $U_{Crit}$ is metallic and above $U_{Crit}$ 
is insulating except at special values of $A/t$, indicated by vertical (dashed) blue lines. At these values of $A/t$, bands of the insulating state touch, leading to a semi-metallic behavior. Panel (b) shows the dependence of the of NN and NNN hopping parameters of the triangular lattice as a function of the driving amplitude $A/t$. The NNN hopping amplitudes ($t_{bb}$, $t_{cc}$) are small, so for visual clarity, 
they are magnified by a factor of 16. In the case shown here, NNN hopping parameter $t_{aa}$ is explicitly chosen to be zero as $t_{ab}=t_{ac}=0.5t$.} 
\label{fig-2}
\end{figure*}
As is usual in slave-rotor decomposition, we rewrite the creation and annihilation operators on the site with interaction as:
\begin{eqnarray}\label{e5}
&&d_{Ia\sigma}^{\dagger}=f_{Ia\sigma}^{\dagger}e^{-i\theta_{Ia}}\ , 
\nonumber \\
&&d_{Ia\sigma}=f_{Ia\sigma}e^{i\theta_{Ia}}\ .
\end{eqnarray}
Here $f^\dagger_{Ia\sigma}$ is the spinon operator at the $a$ site in the $I^{\rm th}$ unit cell and $e^{\pm i\theta_{ia}}$ denotes the rotor creation and annihilation operators defined 
through its action as follows: $e^{\pm i\theta_{I\alpha}}|n_{I\alpha}^{\theta}\rangle=|n_{_{I\alpha}}^{\theta}\pm1\rangle$. At half filling, to preserve the physical Hilbert space on the interacting site we impose the following operator constraint:
\begin{eqnarray}\label{e6}
&&(n_{I a}^{\theta}+n_{I a\uparrow}^{f}+n_{I a\downarrow}^{f})=1\ .
\label{ceq}
\end{eqnarray} 
with electron occupation equal to that of the spinon \ie~$n_{I a\sigma}^{f}=n_{I a\sigma}^{e}$, where $n_{I a\sigma}^{f}=f^\dagger_{Ia\sigma}f_{Ia\sigma}$. We now treat $H^{\rm OR}_{\rm eff}$ within a mean field scheme. We refer the reader to literature for the details of the slave rotor mean field method \cite{florens2004slave, zhao2007self}. Here we briefly outline the method in the context of our calculations. We make a mean field ansatz for the many body ground state\cite{quasi} as $\lvert\Psi\rangle=\lvert\Psi^{fd}\rangle \lvert\Psi^{\theta}\rangle$, where the superscript $d$ refers to a collective index representing the operators on the $b,c$ sub-lattices, the spinon operator on the $a$ sub-lattice are denoted by the  $f$ and the rotor operator by $\theta$ superscript. We note that, considering Hubbard interaction $U$ only on single site is rather interesting as there is spinon contribution emerging from the $a$ site and the free electron contribution from the $b, c$ sites. This type of mixed quasiparticle contribution in the topological edge modes are novel in nature~\cite{sayan2019lieb}. Also this gives rise to correlated metallic behavior as seen in the total density of states of the system~\cite{sayan2019lieb}. Moreover, the scenario of large $U$ on one sublattice and $U=0$ or very small on the other sublattices is not a novel, but routinely occur in strongly correlated materials (for \eg~$\rm CuO_{2}$, $\rm LaNiO_{3}$ etc.)~\cite{keimer2014high, PhysRevLett.110.126404}. The next step is to compute the expectation values $H_{fd} \equiv \langle\Psi^{\theta}\lvert H\rvert \Psi^{\theta}\rangle$ and  $H_{\theta} \equiv\langle \Psi^{fd}|H|\Psi^{fd}\rangle$ which are solved self consistently all the while respecting the above constraint at a mean field level by suitable introductions of chemical potentials. The full detailed expressions are provided in the Appendix~\ref{sec: appendix B}. Here we note that $H_{\theta}$ is the interacting problem that is solved with a cluster mean field theory. For the choice of the hopping amplitudes discussed below, we use a single site rotor cluster, containing one $a$ site, for our results. We have contrasted the results against solution of bigger clusters containing two interacting $a$ sites as well. We now describe the observables calculated before proceeding to the results.
\begin{figure*}[t]
    \centering{
        \includegraphics[width=16.5 cm, height=8.5cm, clip=true]{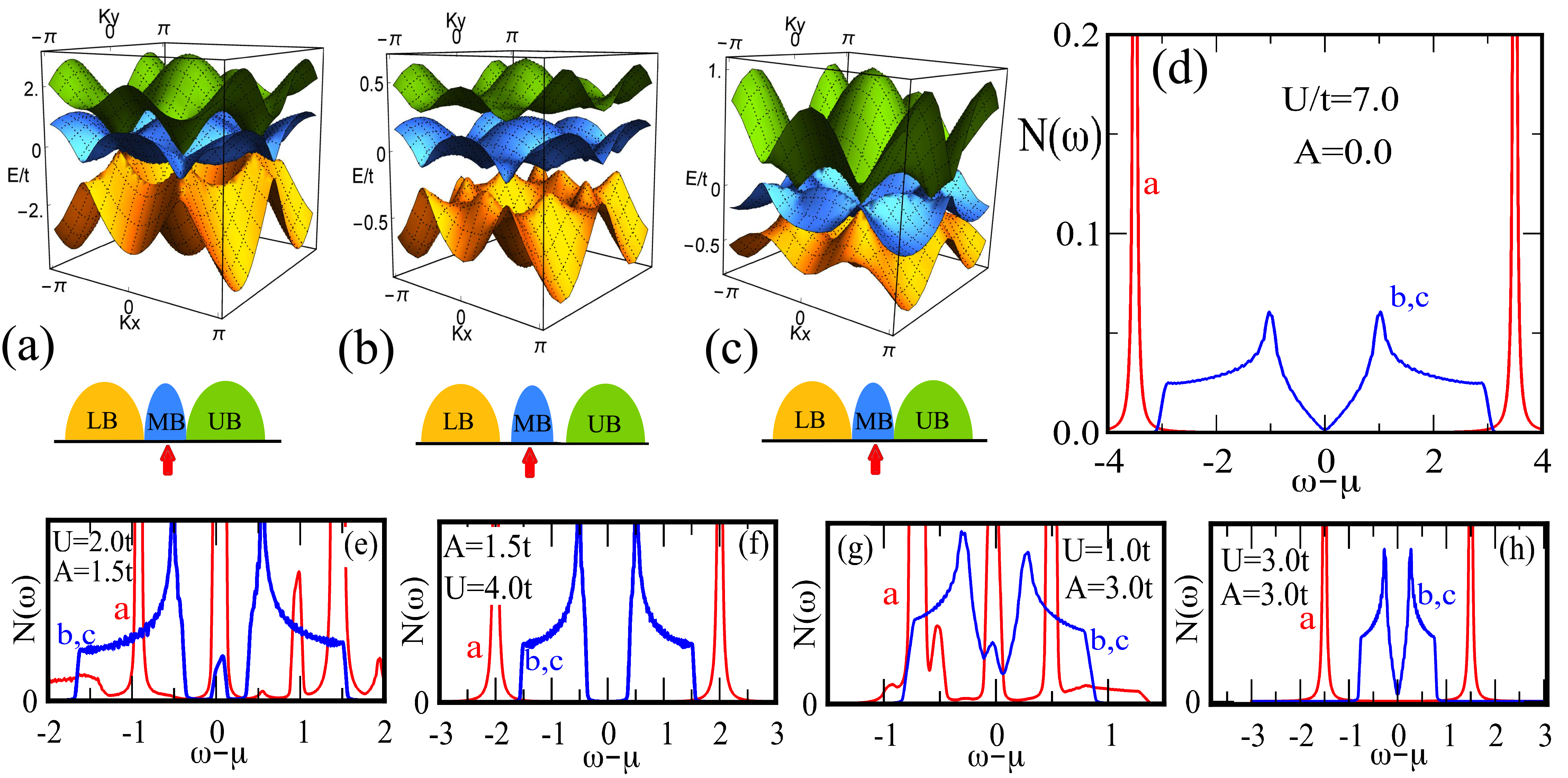}}
    \caption{(Color online) Panels (a) to (c) show the non-interacting tight binding bands of driven triangular lattice. Panel (a) depicts the energy band dispersion and schematic DOS for $A=0$, while panel (b) and (c) manifest the same for $A=1.5t$ and $A=3t$ respectively. In all the three panels, dispersive middle band is half filled which is shown by the small red arrow indicating the position of the chemical potential $\mu$. The band touching points for $A=0$ are removed and band gaps open up at $A\sim 1.5t$. The bands again touch for $A\sim 3t$. Panel (d) shows the sub-lattice resolved DOS for $A=0$ and $U=7t$. Panels (e) to (h), show the evolution of the sub-lattice resolved DOS with $U$, for fixed $A=1.5t$ in (e) and (f) and for $A=3t$ in (g) and (h). 
}
    \label{fig-3}
\end{figure*}

\textit{\underline{Observables}:} Within the SR-MFT, the first important indicator is $\langle\Psi^{\theta}| e^{-i\theta_{Ia}}|\Psi^{\theta}\rangle \equiv\Phi_{Ia}$ and it is assumed to be uniform or site independent. When $\Phi_{Ia}=0$ charge fluctuations on the $a$ sites are suppressed\cite{florens2004slave,zhao2007self}. This signals a `local' Mott transition.  However, with $U$ operative only on the $a$ sub-lattice, \textit{this does not guarantee an insulating ground state}. For the metal-insulator transitions, we  rely on the sub-lattice projected density of states (PDOS) defined as $N_{\gamma}(\omega)=\sum_{\alpha,\sigma}\sum_{i_{\gamma}} 
\lvert\langle\chi_{\alpha}|i_{\gamma},\sigma\rangle\rvert^{2}\delta(\omega-
\epsilon_{\alpha})$, where, $\gamma=a, b, c$ sites in the  $I^{\rm th}$ unit cell. Here, 
$\{|\chi_{\alpha}\rangle\}$ and $\{\epsilon_{\alpha}\}$ correspond to the 
eigenvectors and eigenvalues of $H_{\rm eff}^{\rm OR}$. The derivation is straightforward and is provided in the Appendix~\ref{sec: appendix B}. We also compute the Chern number for the bands and the edge modes in the strip geometry for $H_{fd}$, which provides information of the electron-spinon band topology. There are standard methods to investigate them which are discussed in the Appendix~\ref{sec: appendix C}. For $H_{\theta}$ with more than one $a$ site, as illustrated in Appendix~\ref{sec: appendix D}, we also have to calculate, $\langle\Psi^{\theta}| e^{-i\theta_{Ia}}e^{i\theta_{Ja}}|\Psi^{\theta}\rangle$, which encodes rotor kinetic energy within the enlarged cluster. In the insulating regime, this plays the role of virtual charge fluctuations between the $a$ sites in the rotor cluster, even if $\Phi_{Ia}=\Phi_{Ja}=0$. This will be elaborated more later in the paper.

\section{Results}{\label{sec:III}}
In this section, we present our results for the band spectrum, order parameter and total density of states (DOS). Throughout our analysis, we have chosen $\Omega=10t \gg (t,~U)$, to be in the high frequency regime and $A_{x}=A_{y}=\zeta$, $\phi=\pi/2$ \ie~circularly polarized light as our external electromagnetic drive. We denote the amplitude of the vector potential, for this choice by $A\equiv\sqrt{2}\zeta$.  All energies are measured in units of the non irradiated ($A=0$), bare $b-c$ hopping parameter $t_{bc}\equiv t$, which is chosen to be 1. In these units, we choose $t_{ac}=t_{ab}=0.5t$. From Eq.(\ref{m-so}) we find that for this choice only $b-b$ and $c-c$ NNN hoppings survive in the $1/\Omega$ order term of $H^{\rm OR}_{\rm eff}$. Furthermore, the emergent $t_{bb}$ is always equal to $t_{cc}$. Initially we will focus on results for these choice of parameters, we will then discuss effects on non zero $t_{aa}$ achieved by setting $t_{ab}\neq t_{ac}$. We also use a dimensionless parameter $A/t$ to quantify the magnitude of the vector potential, which is equivalent to measuring the magnetic vector potential in [Ampere $\times $ meter]$^{-1}$. 

Fig.~\ref{fig-2}, shows the $U-A$ phase diagram at zero temperature and summarizes the main results for $t_{ac}=t_{ab}=0.5t$. In panel (b), we plot the variation of the NN and the NNN hopping amplitudes as a function of the radiation amplitude. We begin with the two limiting cases \ie~only $A$ and only $U$, before discussing their combined effects shown in Fig.~\ref{fig-2}(a).
\begin{figure}[t]
    \centering{
        \includegraphics[width=8.5 cm, height=8.5cm, clip=true]{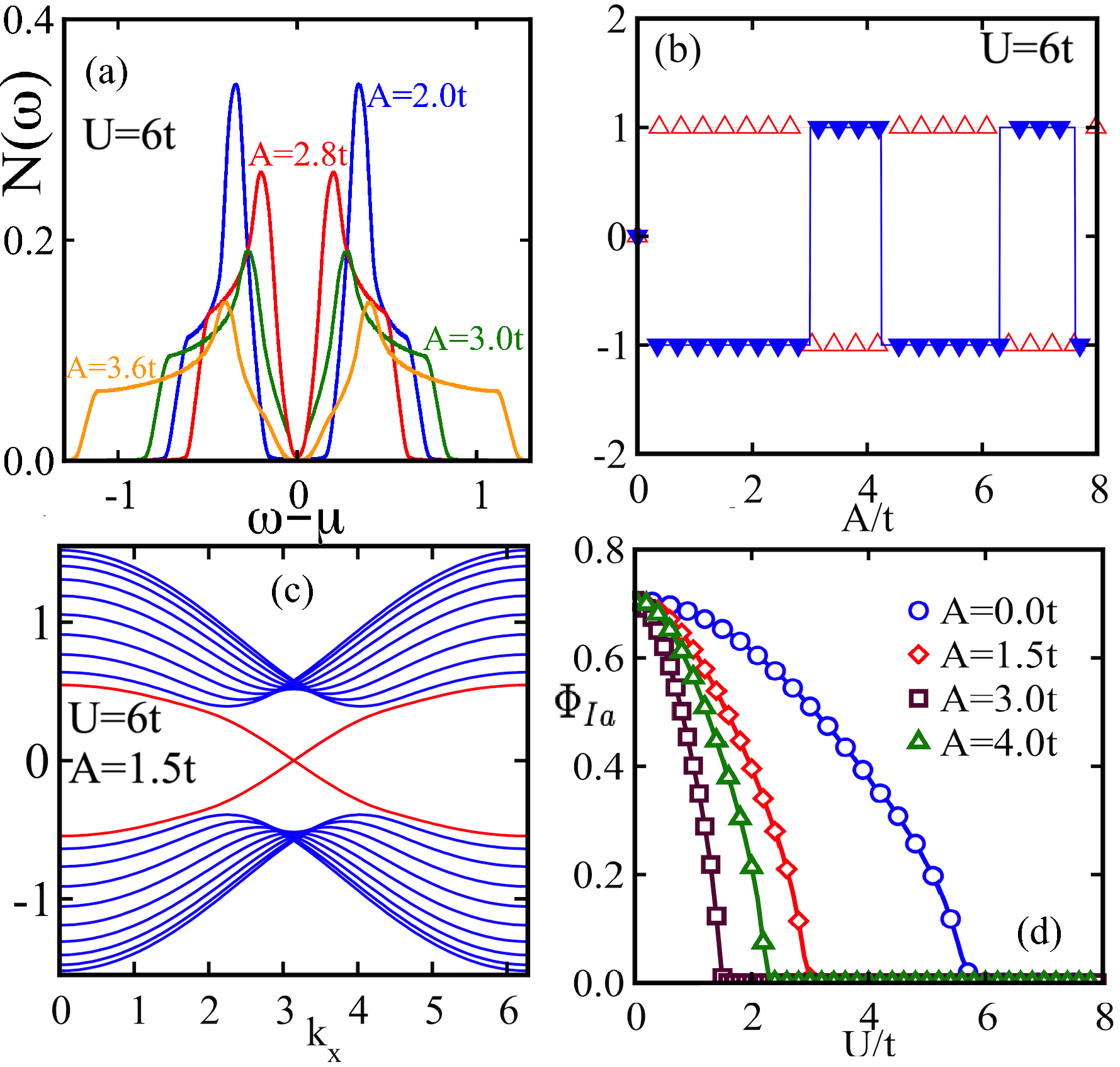}}
    \caption{(Color online)  In panel (a), we show the evolution of the charge gap in the low energy bands as a function of $A$ for $U=6t$. Panel (b) depicts the Chern numbers for the top and the bottom bands of $H_{fd}$ as a function of $A$ for the same $U$(=$6t$), as in panel (a). Panel (c) shows the edge modes for the band calculations in strip geometry for a typical value of $A=1.5t$ in the insulating regime ($U=6t$). The red lines are the topologically protected zero energy edge modes. Panel (d) demonstrates the dependence of order-parameter $\Phi_{Ia}$ on $U$, for various values of $A$. The value of $U$ for local Mott-transition at the $a$ sub-lattice has a non-monotonic dependence on the driving amplitude $A$. 
}
    \vspace{-0.0cm}
    \label{fig-4}
\end{figure}

{\underline{\textit{1. Non-interacting driven system:}}}
For $A=U=0$, the system consists of three dispersive bands, which touch at specific momenta points in the Brillouin zone. The bands are shown in Fig.~\ref{fig-3}(a). From Fig.~\ref{fig-2}(b) we find that the NN hopping amplitudes oscillate with a decaying envelope with increasing $A$, following the Bessel function form in Eq.(\ref{zeroorder}). In addition, the NNN hoping amplitudes  $t_{bb}$ and $t_{cc}$ emerge with $A$ and also oscillate with a decaying envelope. These oscillations are due to the $A$ dependence of the pre-factors $\chi_2$ and $\chi_3$ of $t^2_{ab}-t^2_{bc}$ and $t^2_{ac}-t^2_{bc}$ respectively, which we define as $t_{bb}$ and $t_{cc}$ earlier. The expressions for $\chi_2$ and $\chi_3$ are given in Appendix~\ref{sec: appendix A}, Eq.(\ref{chi}). We find that these oscillations cause a periodic opening and closing of band gap in the non-interacting model. Two typical cases for $A/t\sim 1.5$ and $A/t\sim3.0$, are shown in Fig.~\ref{fig-3}(b)-(c). The schematic of the DOS depicting the closing and opening of the band gaps are shown directly below the bands for the three values of $A$ in panels (a) to (c). The red arrow denotes the chemical potential for half filling. For $A/t\sim$ 1.5, when the bands are separated, we find that the lowest band has a Chern number -1, the middle band is topologically trivial and the highest band has a Chern number +1. From then on, beyond every $A$ at which the bands touch, the upper and lower bands exchange their Chern indices. This shows that at half filling, the driven triangular lattice system always has a metallic ground state for $U=0$, but with non trivial topological energy bands.

{\underline{\textit{2. Large $U$ limit of the purely interacting system:}}}
Fig.~\ref{fig-3}(d) shows the sub-lattice resolved PDOS for $U=7t$ for $A=0$.
We see that large $U$ causes a charge gap, of the order of $U$, in the states from the $a$ sub-lattice, as seen from the red curve. At half filling, the average $a$ site occupation, $\langle n_a\rangle$ is 1. The half filled configuration constitutes the lowest band in the $a$ sub-lattice PDOS, the peak around $\omega-\mu=4$ in panel (d). Any further occupation on the $a$ sub-lattice is pushed up in energy by $U$. The remaining, $b-c$, sub-lattice forms a low energy bands (blue cure) hybridizing through $t_{bc}$. In the large $U$ limit, the effect of the hybridization between the $b-c$ and the $a$ sub-lattices is negligible as virtual charge are suppressed due to large $U$. Thus, the lattice connectivity for the low energy bands are that of a hexagon (see Fig.~\ref{fig-1}(b)) and the system shows graphene like semi-metallic behaviour. From Fig.~\ref{fig-4}(d), we see that $\Phi_{Ia}$ goes to zero for $U>5.8t$ and $A=0$. This corroborates the PDOS by showing that the charge fluctuations are completely suppressed at large $U$ at the $a$ sub-lattice\cite{kondo-mag}. Based on the above discussion, it then implies that for $A=0$ and $U/t>5.8$, the low energy bands, the ones closest to the Femi level, are predominantly constituted by electronic states delocalizing in the $b-c$  sub-lattice, and is a topologically trivial semi-metal.

{\underline{\textit{3. Interaction effects on the driven system:}}}
Fig.~\ref{fig-3}(e) and (f) show the PDOS for $U/t=2$ and $4$, respectively, for fixed $A=1.5t$. We see that all three sub-lattices contribute to the spectral weight at the Fermi energy in (e), while in (f), a finite charge gap is clearly visible. A metallic behavior is also found for small $U/t$ (=1) for $A=3t$ as seen in panel (g), which is similar to panel (e). However on increasing $U$ to $3t$ while staying at $A=3t$, as shown in panel (h), we see a semi-metallic behavior, much like in the $A=0$ case! In $U-A$ phase diagram shown in Fig.~\ref{fig-2}(a), the entire insulating regime has a gapped  DOS as in Fig.~\ref{fig-3} (f), except for special values of $A$ (marked by the vertical blue dashed line) where the semi-metallic ground state, as seen in Fig.~\ref{fig-3}(h), are located. 

To understand this contrasting behavior in the insulating regime, we first note that  as discussed in Sec.~\ref{sec:II}, finite $A/t$ generates NNN $b-b$ and $c-c$ hopping terms that are chiral in nature and thus acts an intrinsic spin-obit coupling. Secondly, in Appendix~\ref{sec: appendix B} we also show that these NNN terms ($t_{bb}$ and $t_{cc}$) are not renormalized by the interaction effects. We find that although $t_{bb}$ and $t_{cc}$ are small in magnitude, with a maximum possible value of $0.018t$ as seen from Fig.~\ref{fig-2}(b) at $A=1.7t$, they have a drastic impact on the bands of the model. Thirdly, in the insulating regime, $\Phi_{Ia}$ is always zero and any contribution from the $a$ sub-lattice are far removed from the Fermi-level, as also seen from the PDOS contributions from the $a$ sub-lattice in Fig.~\ref{fig-3}(f) and (h). Thus, the low energy kinetic energy comes primarily from the $t_{bc}$ with small contributions from NNN $t_{bb}$ and $t_{cc}$.
Thus, the low energy bands (close to the Fermi level) have NN hopping and chiral NNN hopping terms, which is the hopping connectivity of the non interacting Kane-Mele model. Moreover, since the $a$ sites have an average occupation of one electron, the emergent non-interacting Kane-Mele model is also at half filling. 
For a general value of $A$, our calculations reveal that the ground state of this low energy model hosts two distinct bands and that the band gap between these, can be tuned by varying $A$. These low energy bands are shown in Fig.~\ref{fig-4}(a) in the insulating regime ($U=6t$). It shows that the charge gap at $A=2$, reduces and closes at $A=3t$ forming a semi-metal. It then opens again up immediately (data is shown for $A=3.6t$). In addition, the chiral nature of the NNN hopping amplitudes make these low energy bands topologically non-trivial, as is expected in the Kane-Mele model.

Finally, owing to the oscillatory dependence of the hopping amplitudes on $A$, 
the gap closing discussed above, occurs repeatedly for specific values of $A$. 
We compute the Chern number for the $H_{fd}$, after the SR-MFT convergence has been achieved. 
The effect of the rotor Hamiltonian is encoded in the renormalization of the $a-b$ and $a-c$ hopping terms as seen from Eq.(\ref{a2}) in Appendix~\ref{sec: appendix B}. In Eq.(\ref{a2}) we see that $\Phi_{Ia}$ multiplies both $t_{ab}$ and $t_{ac}$, so in the insulating regime ($\Phi_{Ia}=0$) these hopping paths are switched off. Thus calculating the Chern number from the remaining terms in $H_{fd}$ suffices to calculate the topological properties of the low energy Kane-Mele model. Fig.~\ref{fig-4}(b) shows this evolution of the Chern numbers, as a function of $A/t$ for $U/t=6$. As a function of $A$, the Chern numbers of the two bands are swapped periodically. We see that the Chern numbers are swapped exactly at the band touching point $A=3t$, as seen from panel (a). In Fig.~\ref{fig-4}(c) we show the edge states computed from the eigenstates of $H_{fd}$ on a strip geometry for a specific value of $A$ as indicated in the caption. Details of the calculations are presented in Appendix~\ref{sec: appendix C}. We find a linearly dispersing electronic edge mode, as expected for the Kane-Mele model.
\begin{figure}[t]
   \centering{
   \includegraphics[width=8.5 cm, height=4.5cm, clip=true]{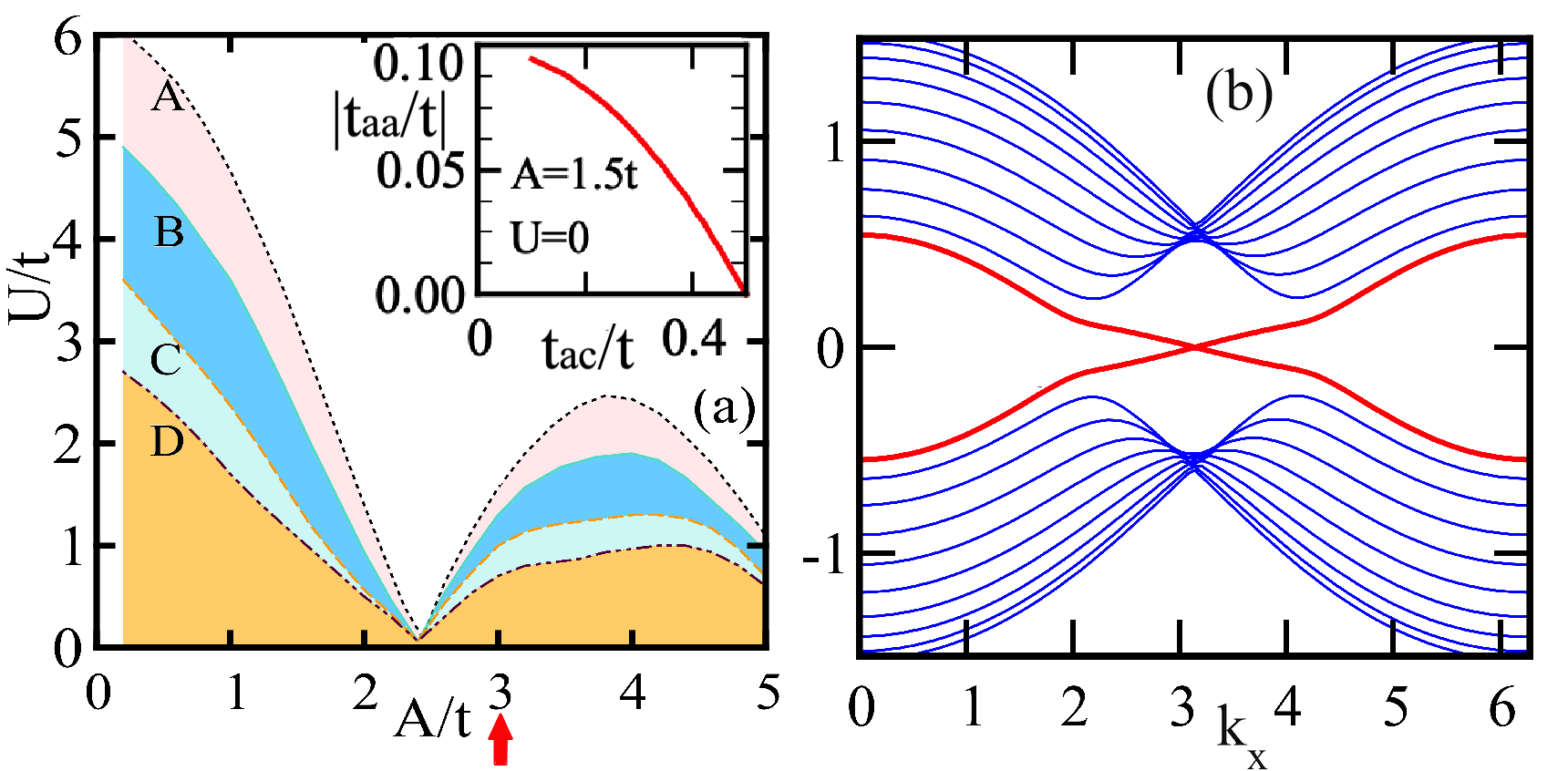}}
   \caption{(Color online) In panel (a) we compare the dependence of $U_{Crit}$ on $A$ for $t_{aa}=0$ (curve A) with that for finite NNN $t_{aa}$. The results are shown for fixed $t_{ab}=0.5t$ and $t_{ac}=$ $0.45t$ (curve B), $0.3t$ (curve C) and $0.2t$ (curve D) which leads to small finite $t_{aa}$. For each $U_{Crit}$ curve, the corresponding region above the curve in an insulator. The vertical red arrow denotes the first band touching value of $A$ in the insulating regime. This location remains unchanged for all the four cases shown.  Inset in panel (a) shows the NNN $|t_{aa}/t|$ as a function of $t_{ac}$ when $U=0$ and $A$ is fixed at $1.5t$. For this $t_{ab}$ is kept fixed at $0.5t$ and $t_{ac}$ is varied between $0.5t$ to $0.1t$. The hopping variation will be discussed in the text. As in Fig.~\ref{fig-2}(b), the NNN hopping amplitude $t_{aa}$ is magnified 16 times. Panel (b) illustrates the edge modes for case B, with $U=6t$ and $A=1.5t$.}
    \label{fig-5}
\end{figure}

We now consider the dependence of  $U_{Crit}$ on $A$, the red (dashed) curve in Fig.~\ref{fig-2}(a). First of all, we observe an overall suppression of $U_{Crit}$ with increasing $A$. This is because the gradual suppression of all the hopping amplitudes with $A$, as seen in panel (b). We also notice that the NN hopping amplitudes oscillate \textit{in-phase} with each other. This modulation of the hopping controls the bandwidth which imprints on the $A$ dependence of $U_{Crit}$, exhibiting that a larger $U$ is needed when the hopping amplitudes are larger. Similarly the minima of $U_{Crit}$ occurs for $A$ values where the three NN amplitudes are closest to zero, effectively narrowing the bandwidth. The value of $U$ minima which $\Phi_{Ia}$ becomes zero also has a similar non-monotonic behavior as seen in panel Fig.~\ref{fig-4}(d). 
For the range of $A$ values shown, the smallest $U$ for $\Phi_{Ia}=0$ occurs for $A=3t$.
This shows that the metal insulator transition is primarily controlled by the competition between various hopping elements and $U$. 

{\underline{\textit{4. Non zero $t_{aa}$ $\&$ spinon contributions:} }}
So far, we have focussed on the case of $t_{ab}=t_{ac}$ for which the emergent NNN $t_{aa}$ always remains zero and have employed a single site rotor cluster in our calculations. A single site rotor cluster cannot capture rotor kinetic energy as there is only one rotor site in $H_{\theta}$. Although, larger clusters are preferred, exponential growth of the rotor Hilbert space forces a compromise. We now use a cluster with four sites, containing two $a$ sites, each connecting to a single $b$ and another $c$ site. The schematic is shown in Appendix~\ref{sec: appendix D}, Fig.~\ref{sfig-1}(b). We have checked that when $t_{aa}=0$, the larger cluster results agree well with the single site cluster results. This is because of the following reason.

First, so far as $H_{fd}$ is concerned, the rotor kinetic energy $\langle\Psi^{\theta}| e^{-i\theta_{Ia}}e^{i\theta_{Ja}}|\Psi^{\theta}\rangle$ only renormalizes $t_{aa}$ as seen from Eq.(\ref{ad1}) in Appendix~\ref{sec: appendix D}. In the insulating regime, even when $\Phi_{Ia}=0$, the rotor kinetic energy term, $\langle\Psi^{\theta}| e^{-i\theta_{Ia}}e^{i\theta_{Ja}}|\Psi^{\theta}\rangle$ remains finite. However, if $t_{aa}=0$ explicitly, there is no change in the Hamiltonian $H_{fd}$. But $H_{\theta}$, now has a new kinetic term as seen in Eq.(\ref{ad2}) which can affect the value of $U_{Crit}$. Nonetheless, particularly in the large $U$ insulating regime, the rotor kinetic energy is suppressed, to the lowest order, by ($\sim |t^{aa}|^2/U$). So the large $U$ band touching are not affected. In our calculation, we find very small renormalization of $U_{Crit}$, but there are no qualitative changes. Below we discuss the case of $t_{aa}\neq 0$, where the corrections to $U_{Crit}$ are significant, but the values of $A$ for the band touchings in the insulating regime, remains unchanged within numerical accuracy.

To perform a systematic study, we have kept $t_{ab}=0.5t$ and varied $t_{ac}$ from $0.5t$ to $0.1t$. The resulting $t_{aa}$, when $U=0$, is shown in the inset of Fig.~\ref{fig-5}(a) for $A=1.5t$, magnified 16 times for better clarity. In Appendix~\ref{sec: appendix D}, Fig.~\ref{sfig-1}(c), we compare the NNN hopping terms, for a range of $A$ values. Although, $t_{aa}$ is of the same order as $t_{bb}$ and $t _{cc}$, as discussed above, $t_{aa}$ is significantly suppressed in the large $U$ insulating limit due to the heavily reduced rotor kinetic term multiplying it. Thus, the band touching locations are unchanged. The $U-A$ phase diagram with many band touching points for the four site cluster calculations and a specific finite $t_{aa}$ are shown in Fig.~\ref{sfig-1}(a) for comparison with Fig.~\ref{fig-2}(a). This shows that the low energy Hamiltonian is, still to a very good approximation, described by a Kane-Mele model. The dispersive edge modes shown in Fig.~\ref{fig-5}(b) still maintain a linear dependence on momentum. The changes in comparison to Fig.~\ref{fig-4}(c) is due to the small rotor kinetic energy that makes the $\langle\Psi^{\theta}| e^{-i\theta_{Ia}}e^{i\theta_{Ja}}|\Psi^{\theta}\rangle\times t_{aa}$ finite. This also shows that the topologically protected edge modes now have both electron and spinon contributions. 
In the main panel of Fig.~\ref{fig-5}, we show $U_{Crit}$ in the $U-A$ plane for four values of $t_{aa}$. The curves are labelled from $A$ to $D$ with increasing magnitude of $t_{aa}$, with $t_{aa}=0$ for curve A. We see that $U_{Crit}$ is progressively suppressed as $t_{aa}$ increases. This in simply due to increase in the bandwidth as additional electron delocalization paths are now available.

{\underline{\textit{5. Experimental feasability:}}}
Triangular lattice has been realized in $\rm ^{87}Rb$, bosonic, cold atom system~\cite{Becker_2010} by the application of three laser beams. Also the Fermi Hubbard model has been engineered in one dimensions ~\cite{jordens2008mott}. In this work ~\cite{jordens2008mott} using fermionic $\rm ^{40}K$ atoms, inter-site tunneling (hopping) amplitude is controlled by the magnitudes of the laser beams, whereas the Hubbard interaction strength can be controlled by changing the occupancy (number of atoms) in the harmonic trap. Further, for periodic driving of bosonic optical system 
($\rm ^{87}Rb$) via sinusoidal shaking of the lattice corresponding to a time-varying linear potential has also been achieved~\cite{PhysRevLett.99.220403}.
Finally, we would like to mention how the spatial modulation of two body interaction in Fermi systems can be realized. The interaction between the fermions is proportional to the scattering length~\cite{spatvar1}, among other quantities. The scattering length is in turn dependent on the coupling strength as well as the energy detuning between the scattering 
and the molecular channels, in the two channel model of Feshbach resonance~\cite{f1}. The coupling between these two channels can be spatially modulated by employing optical Feshbach resonance technique, where counter propagating lasers are applied on the optical lattice. The applied lasers form a spatially varying intensity profile that modulates the interaction strength on the optical lattice. This has been achieved in Fermi systems experimentally~\cite{spatvar0} and analysed theoretically~\cite{spatvar1}.

Based on these, it is clear that the experimental techniques do exist to periodically drive a Fermi system on a triangular lattice with spatially varying interaction strength. Here we briefly mention some typical experimental parameters, gleaned form the above mentioned experiments, that would be relevant for possible realization of our results in cold atomic systems. Typical value of the Hubbard interaction ($U/h$) in fermionic systems studied in one dimensions, are in the range $\sim \rm 5 - 10~KHz$~\cite{jordens2008mott}, where $h$ is the Planck's constant. Also, depending upon the recoil energy and the lattice potential depth one can tune the range of nearest site tunneling from $\rm 0.1~KHz - 1~KHz$~\cite{jordens2008mott, Becker_2010}. Further, for the periodically driven bosonic optical system with $\rm ^{87}Rb$, it has been shown that the amplitude of the effective periodically modulated hopping $t_{eff}(A,\omega)/h$ changes from $\rm 0 - 1~KHz$~\cite{PhysRevLett.99.220403}. Further the frequency of the drive can be within the range $\rm 20 - 30~KHz$ to be in the high frequency regime. To validate our phase diagram, the typical values of the driving strength $A$ and interaction strength $U$ can thus range from $\approx \rm 1 - 10~KHz$ and $\approx \rm 2 - 10~KHz$ respectively in fermionic systems such as in $\rm ^{40}K$ atoms. We emphasize that these parameters are merely a guide for possible experiments.

\section{Summary and Conclusions}{\label{sec:IV}}
To summarize, in this article, we have investigated the impact of strong onsite repulsion in a periodically driven triangular lattice. Driving induces modulation of the bare hopping elements of the triangular lattice and generate chiral NNN hopping terms. In the non-interacting limit, this stabilizes a metal with topologically trivial conduction band and topologically non-trivial, filled and empty bands. Weak interactions on one sub-lattice, for a wide range of bare hopping parameters, leads to repeated metal insulator transitions as a function of the amplitude of the electromagnetic vector potential. In the large interaction  (insulating) regime, the ground state is characterized by widely separated bands emerging from the interacting sub-lattice and small charge gap bands from the rest of the non-interacting sub-lattice. In this limit, the low energy theory (around the Fermi level) is that of a \textit{non-interacting Kane-Mele model that is stabilized by strong interaction effects}, and is shown to host bands with non-trivial topology. We have established that, tuning the amplitude of the electromagnetic vector potential, can be a way to control the intrinsic spin-orbit coupling term of the emergent Kane-Mele model. This can lead to periodic band touchings, that splits the insulating regime by introducing semi-metallic ground states, at which there are topological phase transitions characterized by swapping of band Chern numbers.



\vspace{0.3cm}
\acknowledgements
We acknowledge the use of the SAMKHYA: High Performance Computing Facility provided by IOP, Bhubaneswar and NOETHER cluster at NISER, Bhubaneswar for carrying out our computational work. 
AM acknowledges useful discussions with Kush Saha, NISER, India. We also acknowledge
useful inputs from Ashok Mohapatra, NISER, India.
 
\appendix
\section{Derivation of the effective Floquet Hamiltonian in BW approximation}
\label{sec: appendix A}
In this sub-section we present a full derivation of the effective Hamiltonian used in Eqs.(\ref{bw}-\ref{m-so}). We begin with the tight-binding Hamiltonian of the triangular lattice as written in Eq.(\ref{triangular}) and irradiate it with a laser with vector potential given by $\mathcal{A}(t)=(A_x \cos{\Omega t}, A_y \cos{(\Omega t-\phi)})$. For notational clarity, in this sub-section, the spin indices are suppressed. The hopping amplitudes will pick-up a phase from the Peierl's substitution given by: 
$t\rightarrow t e^{-i\mathcal{A}(t)\cdot\delta_l}$. Here $\delta_l$ is the distance between nearest neighbor sites given by, $\delta_l={\tilde{a}}(\cos{\theta_l}, \sin{\theta_l})$, with 
${\tilde{a}}$ is the lattice spacing, $\theta_l=\frac{\pi}{2}+\frac{2\pi l}{3},\quad l=0,1,2$ as shown in Fig.~\ref{fig-1}(b). This yields,
 \begin{widetext}
 \begin{equation}
  \mathcal{A}(t)\cdot \delta_l={\tilde{a}}A_y\sin{(\phi)}\cos{\left(\frac{2\pi l}{3}\right)}\sin{(\Omega t)}+{\tilde{a}}\left(A_y\cos{\phi}\cos{\left(\frac{2\pi l}{3}\right)}-A_x \sin{\left(\frac{2\pi l}{3}\right)}\right) \cos{\omega t}\ .
 \end{equation}

Using the expression $\exp{\left(-i(r_1\sin{\Omega t}+r_2\cos{\Omega t})\right)}=\sum_{m}J_{-m}\left(\sqrt{r^2_1+r^2_2}\right)\exp{\left[im\left(\Omega t+\arctan{\left(\frac{r_2}{r_1}\right)}\right)\right]}$, 
the time dependent Hamiltonian can be written as, 

 \begin{eqnarray}
H(t)=-\sum_{\langle i,j\rangle}\sum_m J_{-m}(\sqrt{r^2_1+r^2_2})e^{im\left(\Omega t+\arctan{\left(\frac{r_2}{r_1}\right)}\right)} \left(t_{ab}\, a^\dagger_i b_j+ t_{bc} \, b^\dagger_i c_j+
t_{ac} \, a^\dagger_i c_j\right) +h.c\ ,
\label{triangulartime}
 \end{eqnarray}
where $r_1(l)={\tilde{a}}A_y \sin{\phi} \cos{\frac{2\pi l}{3}}$, $r_2(l)={\tilde{a}}\left(A_y \cos{\phi} \cos{\frac{2\pi l}{3}}-A_x \sin{\frac{2\pi l}{3}} \right)$ and $J_m$ is the Bessel function of order $m$. 

The Floquet Hamiltonian is defined as,
\begin{eqnarray}
 H^K=\int^T_0 dt \,H(t)\, e^{iK \Omega t}\ ,
 \label{floquet}
\end{eqnarray} 
 Substituting for Hamiltonian from Eq.(\ref{triangulartime}) in the above equation gives the Floquet Hamiltonian:
  \begin{eqnarray}
H^K=-\sum_{\langle i,j\rangle}\Big[J_{K}\left(\sqrt{r^2_1+r^2_2}\right)e^{-iK\arctan{\left(\frac{r_2}{r_1}\right)}} \left(t_{ab}\, a^\dagger_i b_j+ t_{bc} \, b^\dagger_i c_j+
t_{ac} \, c^\dagger_i a_j\right) \nonumber\\
+J_{-K}\left(\sqrt{r^2_1+r^2_2}\right)e^{-iK\arctan{\left(\frac{r_2}{r_1}\right)}} \left(t_{ab}\, b^\dagger_i a_j+ t_{bc} \, c^\dagger_i b_j+
t_{ac} \, a^\dagger_i c_j\right)\Big]\ .
\label{triangularf}
\end{eqnarray}

In this paper, we consider a high-frequency limit where the frequency of the drive is larger than the band width of the system. As mentioned in Sec.~\ref{sec:II}, only virtual photon transitions 
are allowed in this limit. Therefore one can find an effective quasi-static Hamiltonian using one of high-frequency expansion schemes such as  Brillouin-Wigner~\cite{mikami2016brillouin}, Floquet-Magnus~\cite{mananga2011introduction,blanes2009magnus} or van Vleck~\cite{eckardt2015high,bukov2015universal}. We use the 
Brillouin-Wigner scheme where the effective Hamiltonian to order $O(1/\Omega)$ has the form,
\begin{eqnarray}
 H_{K}=H^0+ \sum_{n\ne 0}\frac{H^{-n} H^{n}}{n\Omega}\nonumber\ ,
 \end{eqnarray}
The zeroth order term is given by,
\begin{eqnarray}
 H^0= -\sum_{\langle ij \rangle} J_{0}\left(\sqrt{r^2_1+r^2_2}\right) \left( t_{ab}\,  a^\dagger_i b_j + t_{bc}\,  b^\dagger_i c_j+t_{ac}\,  c^\dagger_i a_j +h.c.\right)\ , \nonumber
\end{eqnarray}
and the $O(1/\Omega)$ terms are given by,
\begin{eqnarray}
 \frac{H^{-n} H^{n}}{n\omega}&=&
 \sum^{\mathcal{P}_1}_{\langle \langle ij \rangle\rangle} \chi_1 \, \nu_{ij} \, \left( (t^2_{ab}-t^2_{ac})   a^\dagger_i a_{j} 
+(t^2_{ab}-t^2_{bc})  b^\dagger_i b_{j}  +(t^2_{ac}-t^2_{bc})  c^\dagger_i c_{j} \right) \nonumber\\
&+&\sum^{\mathcal{P}_2}_{\langle \langle ij \rangle\rangle} \chi_2 \, \nu_{ij}  \,\left( (t^2_{ab}-t^2_{ac})   a^\dagger_i a_{j} 
+(t^2_{ab}-t^2_{bc})  b^\dagger_i b_{j}  +(t^2_{ac}-t^2_{bc})  c^\dagger_i c_{j} \right)  \nonumber\\
 &+&\sum^{\mathcal{P}_3}_{\langle \langle ij \rangle\rangle}\chi_3 \, \nu_{ij}  \, \left( (t^2_{ab}-t^2_{ac})   a^\dagger_i a_{j} 
+(t^2_{ab}-t^2_{bc})  b^\dagger_i b_{j}  +(t^2_{ac}-t^2_{bc})  c^\dagger_i c_{j} \right).
\label{so}
 \end{eqnarray}
where,  
\begin{eqnarray}
 \chi_1&=&\frac{i}{n\omega} J_n({\tilde{a}}A_y) J_n\left(\frac{{\tilde{a}}}{2}\sqrt{A^2_y+3A^2_x+2 \sqrt{3}A_x A_y\cos{\phi}}\right)
 \sin{\left(n\left[\frac{\pi}{2}-\phi-\arctan\left(\frac{A_y\cos{\phi}+\sqrt{3}A_x}{A_y\sin{\phi}}\right)\right]\right)}\nonumber\\
 \chi_2 &=&\frac{i}{n\omega}J_n({\tilde{a}}A_y) J_n\left(\frac{{\tilde{a}}}{2}\sqrt{A^2_y+3A^2_x-2 \sqrt{3}A_x A_y\cos{\phi}}\right)
 \sin{\left(n\left[\frac{\pi}{2}-\phi-\arctan\left(\frac{A_y\cos{\phi}-\sqrt{3}A_x}{A_y\sin{\phi}}\right)\right]\right)} \nonumber\\
 \chi_3 &=&\frac{i}{n\omega} J_n\left(\frac{{\tilde{a}}}{2}\sqrt{A^2_y+3A^2_x+2 \sqrt{3}A_x A_y\cos{\phi}}\right) J_n\left(\frac{{\tilde{a}}}{2}\sqrt{A^2_y+3A^2_x-2 \sqrt{3}A_x A_y\cos{\phi}}\right)\nonumber\\
 && \sin{\left(n\left[\arctan\left(\frac{A_y\cos{\phi}+\sqrt{3}A_x}{A_y\sin{\phi}}\right)-\arctan\left(\frac{A_y\cos{\phi}-\sqrt{3}A_x}{A_y\sin{\phi}}\right)\right]\right)}  
 \label{chi}
\end{eqnarray}
We are interested in the case of circular polarization where $A_x=A_y$ and $\phi=\frac{\pi}{2}$. The values of $\chi_{1,2,3}$ for circular polarization are given by,
$\chi_1=\chi_2=\chi_3=\frac{i J^2_n({\tilde{a}}A)  \sin{\frac{2\pi n}{3}}}{n\omega}$.

\section{Slave-rotor mean field treatment of the off-resonant Hamiltonian}
\label{sec: appendix B}
Within single site cluster mean field theory we can write the spinon-electron Hamiltonian $H_{fd}=\langle\Psi^{\theta}|H_{\rm eff}^{\rm OR}|\Psi^{\theta}\rangle$ and rotor Hamiltonian $H_{\theta}=\langle\Psi^{fd}|H_{\rm eff}^{\rm OR}|\Psi^{fd}\rangle$ as,
\begin{eqnarray}\label{a2}
H_{fd}&=&\sum_{ 
    I,J,\beta,\sigma}(\langle\Psi^{\theta}| 
e^{-i\theta_{Ia}}|\Psi^{\theta}\rangle
t^{A,NN}_{Ia\sigma;J\beta\sigma}
f_{Ia\sigma}^{\dagger}d_{J\beta \sigma} +
h.c.)+\sum_{I,J,\sigma}(t^{A,NN}_{Ib\sigma;Jc\sigma}
d_{Ib\sigma}^{\dagger}d_{Jc\sigma} +
h.c.)\\\nonumber &+&\sum_{ 
    I,J,\alpha,\sigma}(t^{A,NNN}_{I\alpha\sigma;J\alpha\sigma}d_{I\alpha\sigma}^{\dagger}d_{J\alpha\sigma}+h.c.)+ U/2\sum_{I}\langle\Psi^{\theta}|
n_{Ia}^{\theta}(n_{Ia}^{\theta}-1)|\Psi^{\theta}\rangle-\mu_fN_f\ \ 
\end{eqnarray}
\begin{eqnarray}\label{a3}
H_{\theta}&=&\sum_{ 
    I,J,\beta,\sigma}( t^{A,NN}_{Ia\sigma;J\beta\sigma}
\langle\Psi^{fd}|f_{Ia\sigma}^{\dagger}d_{J\beta \sigma}|\Psi^{fd}\rangle e^{-i\theta_{Ia}}+
h.c.)+\sum_{ 
    I,J,\sigma}\langle\Psi^{fd}|(t^{A,NN}_{Ib\sigma;Jc\sigma}
d_{Ib\sigma}^{\dagger}d_{Jc\sigma} +
h.c.)|\Psi^{fd}\rangle\\\nonumber &+&\sum_{ 
    I,J,\alpha,\sigma}\langle\Psi^{fd}| (t^{A,NNN}_{I\alpha\sigma;J\alpha\sigma}d_{I\alpha\sigma}^{\dagger}d_{J\alpha\sigma}+h.c.)|\Psi^{fd}\rangle+U/2\sum_{I}
    n_{Ia}^{\theta}(n_{I,a}^{\theta}-1)-\mu_\theta 
    N_\theta\ \ 
\end{eqnarray}  
Here, $t^{ANN}_{I\alpha\sigma;J\beta\sigma}$ and $t^{ANNN}_{I\alpha\sigma;J\alpha\sigma}$ are the NN and NNN hopping amplitudes, that were contained in $t^{A}_{I\alpha\sigma;J\beta\sigma}$ in $H_{\rm eff}^{\rm OR}$, in the main paper. In the first term of Eq.(\ref{a2}), $\beta$ runs over $b$ and $c$. This term is the $a-b$ and $a-c$, NN kinetic energy terms, where the 
bare $t_{ab}$ and $t_{ac}$ are renormalized by $\Phi_{Ia}=\langle\Psi^{\theta}| 
e^{-i\theta_{Ia}}|\Psi^{\theta}\rangle$. Within the mean field ansatz, $\Phi_{Ia}$ is assumed to be same on all $a$ sites. The second term is the bare kinetic term containing only $b-c$ hopping. In the third term of Eq.(\ref{a2}), the $\alpha$ summation goes over $b$ and $c$ and contains only $t_{bb}$ and $t_{cc}$ NNN hopping terms. This in because, the emergent NNN hopping amplitude $t_{aa}$ is zero for our choice on bare hopping amplitudes \ie~$t_{ab}=t_{ac}$. 
We note that in $H_{fd}$, $\Phi_{Ia}$, which is assumed to be same on each $a$ site, scales the NN hopping terms only. Thus when the local Mott transition occurs, the NN term in the Hamiltonian vanishes.

We start with a given $\Phi_{Ia}$, assumed to be same on all $a$ sites within the single site rotor cluster mean field theory, and then diagonalize the spinon Hamiltonian. Now with the resultant eigenvectors 
and eigenvalues average $\Phi_{Ia}$ is calculated. The constraint equation, Eq.(\ref{ceq}) in the main paper, is imposed at a mean field level, which reads as:
\begin{eqnarray}\label{e6}
&&\langle n_{I a}^{\theta}\rangle +\langle n_{I a\uparrow}^{f}\rangle +\langle n_{I a\downarrow}^{f}\rangle =1\ 
\end{eqnarray}
The  average spinon occupation on the site $a$ is identified to the electron occupation and the chemical potential $\mu_{\theta}$ is adjusted to fix the rotor occupation so that it satisfies the mean field constraint equation.
In our calculation, the Hilbert space for the interacting rotor Hamiltonian, is restricted by limiting the local $a$ site occupation to a maximum occupation of 3. Then  $\Phi_{Ia}$ is used in $H_{fd}$ and 
the spinon Hamiltonian is rediagonalized. The process is repeated till energy convergence is achieved.\\

{\bf{\textit{\underline{DOS calculation:}}}}
The main observable we focus on is
the sub-lattice projected density of states (PDOS). The PDOS is defined in general as,
$N_{\gamma}(\omega)=\sum_{\alpha,\sigma}\sum_{i_{\gamma}} 
\lvert\langle\chi_{\alpha}|i_{\gamma},\sigma\rangle\rvert^{2}\delta(\omega-
\epsilon_{\alpha})$, where, $\gamma=a, b, c$ sites in the  $I^{\rm th}$ unit cell. Here, 
$\{|\chi_{\alpha}\rangle\}$ and $\{\epsilon_{\alpha}\}$ correspond to the 
eigenvectors and eigenvalues of $H$. PDOS calculation is standard and is not repeated here. Below we focus on the PDOS for the interacting sub-lattice $a$.
Since we have split the electron 
into a rotor and a spinon at every $a$ site of our problem, we first need to reconstruct 
the (electron) single particle Green's function and then take its imaginary part to 
compute the spectral function and the PDOS. To do so, we begin with the local 
(on-site) retarded Matsubara Green's function at a site $a$ in a unit cell $I$, which can be defined as,
\begin{eqnarray}\label{e10}
G_{Ia\sigma}(i\omega_{n})&=&
-\int_{0}^{\beta} d\tau  e^{i\omega_{n}\tau} 
\langle \Psi|
d_{Ia,\sigma}(\tau)d_{Ia,\sigma}^{\dagger}(0)|\Psi\rangle \\\nonumber 
&=&-\int_{0}^{\beta}d\tau e^{i\omega_{n}\tau}\langle \Psi^{fd} |
f_{I\alpha\sigma}(\tau)f_{I\alpha\sigma}^{\dagger}(0)|\Psi^{fd}\rangle \\\nonumber
&&~~~~~~~~~~~~~~~\times \langle \Psi^\theta| e^{-i\theta_{I\alpha}(\tau)} 
e^{i\theta_{I\alpha}(0)}|\Psi^\theta\rangle\ .
\end{eqnarray}
The above decomposition of electron Green's function into a convolution of rotor 
and spinon Green's functions is possible for the chosen mean field ansatz 
$|\Psi^{fd}\rangle |\Psi^{\theta}\rangle$.
The spinon correlator in Eq.(\ref{e10}) can be calculated as
\begin{eqnarray}\label{e11}
&&\frac{1}{2}\sum_{\sigma}\langle 
f_{I\alpha\sigma}(\tau)f_{I\alpha\sigma}^{\dagger}(0)\rangle \nonumber \\
&&=\frac{1}{2}\sum_{\alpha\sigma}\lvert \langle 
\chi^f_{\alpha}|I\alpha,\sigma\rangle 
\rvert 
^{2} [1-n_{f}(\epsilon^f_{\alpha}-\mu_{f})]e^{-\tau(\epsilon^f_{\alpha}-\mu_{f})}\ .
\label{SC}
\end{eqnarray}
Here, $\{|\chi^f_\alpha\rangle\}$ and $\{\epsilon_{\alpha}^f\}$ are the spinon 
eigenvectors and eigenvalues respectively. The rotor correlator in Eq.(\ref{e10}) 
can be expressed as 
\begin{eqnarray}\label{e12}
&&\langle e^{-i\theta_{I\alpha,\sigma}(\tau)}e^{i\theta_{I\alpha,\sigma}(0)}\rangle 
\nonumber \\
&&= \frac{1}{Z_{\theta}}\sum_{m,n}e^{-\beta\epsilon_{m}}\langle m\lvert 
e^{-i\theta_{I\alpha,\sigma}}\rvert n \rangle \langle n \rvert 
e^{i\theta_{I\alpha,\sigma}} \rvert m\rangle 
e^{\tau(\epsilon_{m}-\epsilon_{n})}\ .
\end{eqnarray}
where, $\{\epsilon_{m}\}$ and $\{\rvert m\rangle\} $ are the eigenvalues and 
corresponding eigenvectors of the rotor Hamiltonian $H_{\theta}$. 
Here, $Z_{\theta}$ is the rotor partition function defined as 
$\sum_{m}e^{-\beta\epsilon_{m}}$. Using Eq.(\ref{e10}), the integration over 
imaginary time $\tau$ can be performed. We then analytically continue back to 
the real frequency to obtain $G_{Ia\sigma}(\omega)$. The PDOS is obtained 
from the corresponding imaginary part as usual.

\section{Chern numbers $\&$ edge mode calculations}
\label{sec: appendix C}
The Chern number is defined for n$^{\rm th}$ Bloch band as, 
\begin{eqnarray}
C_{n}&=&
 \frac{1}{2\pi i}\int_{T}^{}
d^{2}k F_{ij}(k)\ ,
\label{so}
\end{eqnarray}
The integration is over the full two dimensional Brillouin zone. The Berry connection and associated curvature is defined as, 
\begin{eqnarray}
A_{i}(k)&=&
 <n(k)|\partial_{i}|n(k)>\ ,
\label{so}
\end{eqnarray}

\begin{eqnarray}
F_{i,j}&=&
 \partial_{i}A_{j}-\partial_{j}A_{i}\ ,
\label{so}
\end{eqnarray}
where $n(k)$ is the normalised Bloch wave function of the non degenerate n$^{\rm th}$ Bloch band that is calculated by diagonalizing the momentum space Bloch Hamiltonian. For numerical calculation we employ the method discussed in the reference~\cite{takahiro2013chern}.

To obtain a solution for the edge state, we have to consider a ribbon geometry. Here we adopt the periodic boundary condition along $x$ axis and along $y$ we have an open boundary condition. Thus, $k_{x}$ is a good quantum number here. Now performing the Fourier transformation only along the $x$ direction, the problem is reduced to one dimensional (1D) having only $k_{x}$ as the variable. Hence, the 1D Hamiltonian can be written as,
\begin{eqnarray}
 H_{k_{x}}&=&
 \sum^{N}_{ 
I,J,\alpha,\beta,\sigma} t^{A,k_{x}}_{I\alpha\sigma;J\beta\sigma}
d_{I\alpha\sigma}^{k_{x}\dagger}d^{k_{x}}_{J\beta \sigma} +h.c. \ ,
\label{so}
\end{eqnarray}
where, $I$, $J$ denote the site indices, $N$ is the total number of lattice sites along $y$ direction, $\alpha,\beta$ run oven $a$, $b$ and $c$. $t^{A,k_{x}}_{I\alpha\sigma;J\beta\sigma^{\prime}}$ contains all the nearest and next nearest Fourier transformed hopping terms. Now diagonalizing the Hamiltonian numerically we obtain the required 1D edge spectra. In our case, for the edge calculation we consider
$A=1.5t$ and $N=10$. Both the Chern number and the edge mode calculations are carried out for $H_{fd}$, once the SR-MFT results have converged. 
\begin{figure*}[t]
	\centering{
		\includegraphics[width=18.0cm, height=6cm, clip=true]{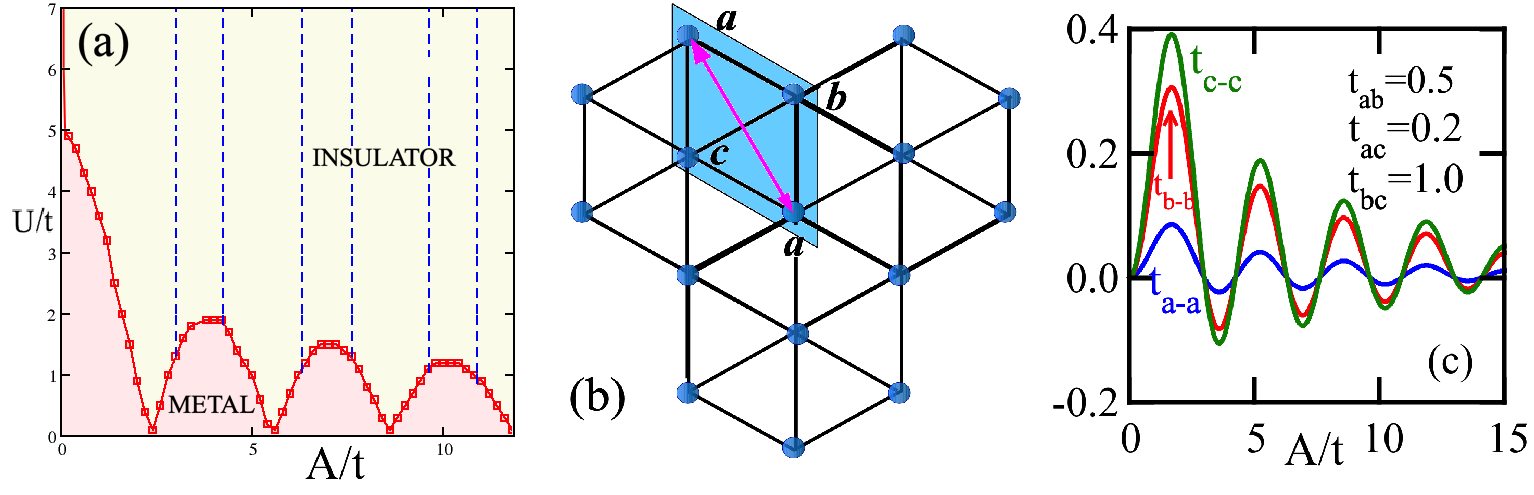}}
	\caption{(Color online) $U-A$ phase diagram for non zero $t_{aa}$, when $t_{ac}=0.45t$ and $t_{ab}=0.5t$, is shown in panel (a). Panel (b) depicts the schematic of the lattice, where the four site cluster (shaded parallelogram) is used in the rotor calculation. The NNN hopping element $t_{aa}$ within the cluster is shown by the magenta arrow. (c) The NNN hopping amplitudes as a function of $A$ are shown for the chosen bare NN hopping parameters as indicated.
	 }
	\label{sfig-1}
\end{figure*}
\section{Two $a$ site in the rotor cluster}
\label{sec: appendix D}
In the previous single site cluster calculation we take nearest neighbour hopping in such a way that light induced $a$-$a$ next-nearest neighbour is vanished. Now we allow small NNN $a$-$a$ hopping by considering $t_{ab}\neq t_{ac}$. The corresponding spinon-electron and rotor Hamiltonians read as,
\begin{eqnarray}\label{ad1}
H_{fd}&\rightarrow H^{\square}_{fd}=H_{fd}+\sum_{ 
    I,J,\sigma}(\langle\Psi^{\theta}| e^{-i\theta_{Ia}}e^{i\theta_{Ja}}|\Psi^{\theta}\rangle t^{A,NNN}_{Ia\sigma;Ja\sigma}f_{Ia\sigma}^{\dagger}f_{Ja\sigma}+h.c.)\\\nonumber
\end{eqnarray}
\begin{eqnarray}\label{ad2}
H_{\theta}\rightarrow H^{\square}_{\theta}=H_{\theta}+\sum_{ 
    I,J,\sigma}(\langle\Psi^{fd}|f_{Ia\sigma}^{\dagger}f_{Ja\sigma}|\Psi^{fd}\rangle t^{A,NNN}_{Ia\sigma;Ja\sigma} e^{-i\theta_{Ia}}e^{i\theta_{Ja}}+h.c.) \\\nonumber
\end{eqnarray}  

Here, $t^{A,NNN}_{Ia\sigma;Ja\sigma^{\prime}}$ denotes the light driven direct $a-a$ hopping amplitudes. This additional term, in $H^{\square}_{\theta}$ is further decoupled in a kinetic term decoupling scheme $e^{-i\theta_{Ia}}e^{i\theta_{Ja}}\rightarrow \langle\Psi^{\theta}|e^{-i\theta_{Ia}}|\Psi^{\theta}\rangle e^{i\theta_{Ja}}$ or $\Phi_{Ia}e^{i\theta_{Ja}}$. Since $\Phi_{Ia}$ is assumed to be uniform, this implies that the additional term in $H^{\square}_{\theta}$ reduces to a local term. The solution of the four site cluster, then proceeds exactly like the single site cluster case discussed in Appendix~\ref{sec: appendix B}. The difference being that a four site rotor Hamiltonian is diagonalized instead of a single site cluster. The four site ($a$-$b$-$a$-$c$) parallelogram cluster, indicated by the shaded parallelogram (enclosed within the dashed lines) is shown in Fig.~\ref{sfig-1}(b). Also, as for the one site cluster, we allow local rotor occupation of 3 ion each $a$ site in $H^{\square}_{\theta}$.

We note that in addition to $\langle\Psi^{\theta}| 
e^{-i\theta_{Ia}}|\Psi^{\theta}\rangle$ or $\Phi_{Ia}$ that renormalizes the NN hopping, $\langle\Psi^{\theta}| e^{-i\theta_{Ia}}e^{i\theta_{Ja}}|\Psi^{\theta}\rangle $ now scales the NNN hopping between the $a$ sites. In the insulating phase when $\Phi_{Ia}=0$, the rotor kinetic energy within the cluster $\langle\Psi^{\theta}| e^{-i\theta_{Ia}}e^{i\theta_{Ja}}|\Psi^{\theta}\rangle $ is not zero. It encodes the virtual charge fluctuations. In the insulating phase, the  NN $a-b$ and $a-c$ hopping amplitudes are renormalized to zero (as $\Phi_{Ia}=0$). In this regime NN $t_{bc}$, the NNN $a-a$, $b-b$ and $c-c$ terms survive. When $t_{aa}$ is chosen to be zero $H^{\square}_{fd}$ has the hopping connectivity of the Kane-Mele model. When $t_{aa}\neq 0$, the virtual charge fluctuations within the four site rotor cluster allow a $a-a$ hopping mediated spinon kinetic energy contribution to the total kinetic energy in $H^{\square}_{fd}$. The resultant phase digram is shown in Fig.~\ref{sfig-1}(a). In Fig.~\ref{fig-5}(a), only the first band touching point was indicated by the red arrow. Here for completeness, for a specific value of non-zero $t_{aa}$, we show all the band touching points in the phase diagram (see Fig.~\ref{sfig-1}(a)) as was shown for the $t_{aa}=0$ case in Fig.~\ref{fig-2}(a). Comparison shows that these remain qualitatively unchanged, as is discussed in the subsection 4 in the main paper. In Fig.~\ref{sfig-1}(c), the variation of 
all NNN hopping parameters are shown as a function of $A$ for the bare parameters indicated in the figure.

\end{widetext}
 
\bibliography{bibfile}{}

\begin{thebibliography}{87}
\bibitem [{qua()}]{quasi}%
  \BibitemOpen
  \href@noop {} {}\bibinfo {note} {Strictly speaking all bands in the
  off-resonant driven system are quasi-energy bands in the Floquet spectrum.
  However for brevity we will simply refer to them as bands in the paper.
  Similarly, the ground states mentioned in the paper are lowest energy states
  in the quasi energy spectra.}\BibitemShut {Stop}%
\bibitem [{kon()}]{kondo-mag}%
  \BibitemOpen
  \href@noop {} {}\bibinfo {note} {We note that the trapped charges at the $a$
  sub-lattice form quantum local moments. In the present paper however our
  modest aim is to focus on the physics of charge fluctuations and do not
  consider the possibility of Kondo screening or RKKY mediated magnetic
  order.}\BibitemShut {Stop}%
\end{thebibliography}

\end{document}